\documentstyle[11pt,aaspptwo,flushrt,tighten]{article} 

\newcommand{\ltsima} {$\; \buildrel < \over \sim \;$}
\newcommand{\gtsima} {$\; \buildrel > \over \sim \;$}
\newcommand{\lta} {\lower.5ex\hbox{\ltsima}}
\newcommand{\gta} {\lower.5ex\hbox{\gtsima}}
\slugcomment{ACCEPTED FOR PUBLICATION IN THE ASTROPHYSICAL JOURNAL}

\begin{document}

\title{ASCA OBSERVATIONS OF THE GALACTIC BULGE \\ 
HARD X--RAY SOURCE GRS 1758--258} 

\author{S. Mereghetti\altaffilmark{1}, D. I. Cremonesi\altaffilmark{2}, 
F. Haardt\altaffilmark{3}, T. Murakami\altaffilmark{4},
T. Belloni\altaffilmark{5} \& A. Goldwurm\altaffilmark{6}} 

\altaffiltext{1}{Istituto di Fisica Cosmica del CNR, via Bassini 15, I-20133
Milano, Italy, sandro@ifctr.mi.cnr.it,}

\altaffiltext{2}{Dipartimento di Fisica, Universit\`a di Milano, via Celoria 16, 
I-20133 Milano, Italy, davide@ifctr.mi.cnr.it,}

\altaffiltext{3}{Department of Astronomy \& Astrophysics, Institute of
Theoretical Physics, G\"oteborg University \& Chalmers University
of Technology, 412 96 G\"oteborg, Sweden, haardt@fy.chalmers.se,}

\altaffiltext{4}{Institute of Space and Astronautical Science, 3-1-1, 
Yoshinodai, Sagamihara, Kanagawa 229, Japan, 
murakami@astro.isas.ac.jp,}

\altaffiltext{5}{Astronomical Institute "Anton Pannekoek", University of 
Amsterdam, Kruislaan 403, 1098 SJ Amsterdam, The 
Netherlands, tmb@astro.uva.nl,}

\altaffiltext{6}{Service d'Astrophysique, CEA, CEN Saclay, 91191 Gif-sur-
Yvette Cedex, France, goldwurm@sapvxg.saclay.cea.fr.}

\begin{abstract} 
GRS 1758--258 is one of the few persistent hard X--ray 
emitters ($E>100$ keV) in the Galaxy. Using the ASCA satellite, 
we have obtained the first detailed data on GRS 1758--258 in the 
1--10 keV range, where previous observations were affected by 
confusion problems caused by the nearby strong source GX5--1.  
The spectrum is well described by a power law with photon 
index 1.7 without strong Fe emission lines.  A prominent soft 
excess, as observed with ROSAT when the hard X--ray flux was 
in a lower intensity state, was not detected. However, the 
presence of a soft spectral component, accounting for at most 5\% 
of the 0.1--300 keV flux, cannot be excluded. The accurate 
measurement of interstellar absorption 
($N_H=(1.5\pm 0.1)\times 10^{22}$ cm $^{-2}$) 
corresponds to an optical extinction which definitely 
excludes the presence of a massive companion.   
\end{abstract}


\twocolumn

\section{Introduction}

The source GRS 1758--258 was discovered in the hard X--ray/soft 
$\gamma$--ray energy range with the SIGMA/GRANAT coded mask 
telescope (Sunyaev et al. 1991). GRS 1758--258 is of particular 
interest since, together with the more famous source 1E 1740.7--
2942, it is the only persistent hard X--ray emitter ($E>100$ keV) in 
the vicinity of the Galactic Center (Goldwurm et al. 1994). Both 
sources have peculiar radio counterparts with relativistic jets 
(Mirabel et al. 1992a; Rodriguez, Mirabel \& Mart\`i 1992; Mirabel 
1994) and might be related to the 511 keV line observed from 
the Galactic Center direction (Bouchet et al. 1991). Despite the 
precise localization obtained at radio wavelengths, an optical 
counterpart of GRS 1758--258 has not been identified 
(Mereghetti et al. 1992; Mereghetti, Belloni \& Goldwurm 1994a). 
Simultaneous ROSAT and SIGMA observations, obtained in the 
Spring of 1993, indicated the presence of a soft excess 
(Mereghetti, Belloni \& Goldwurm 1994b). This spectral 
component ($E<2$ keV) was weaker in 1990, when the hard X--ray 
flux ($E>40$ keV) was in its highest observed state. On the basis of 
its hard X--ray spectrum, GRS 1758--258 is generally considered 
a black hole candidate (Tanaka \& Lewin 1995; Stella et al. 
1995). The possible evidence for a soft spectral component 
anticorrelated with the intensity of the hard ($>40$ keV) emission 
supports this interpretation. 
No detailed studies of GRS 1758--258 in the "classical" X--ray 
range have been performed so far. Here we report the first 
observations of this source obtained with an imaging instrument 
in the $0.5-10$ keV energy range. 

\section{Data Analysis and Results}

The observation of GRS 1758--258 took place between 1995 
March 29 22:39 UT and March 30 15:38 UT.  The ASCA satellite 
(Tanaka, Inoue \& Holt 1994) provides simultaneuos data in four 
coaligned telescopes, equipped with two solid state detectors 
(SIS0 and SIS1) and two gas scintillation proportional counters 
(GIS2 and GIS3). 
We applied stringent screening criteria to reject periods of high 
background, and eliminated all the time intervals with the 
bright earth within 40 degrees of the pointing direction for the 
SIS data (10 degrees for the GIS), resulting in the net exposure 
times given in Table 1.  

\begin{tabular}{|l|c|c|}
\hline
\multicolumn{3}{|c|}{TABLE 1}\\
\hline
\hline
&Exposure Time (s)&Count Rate (counts/s)\\
SIS0&9,471&5.310$\pm$0.024\\
SIS1&9,455&4.220$\pm$0.022\\
GIS2&12,717&4.507$\pm$0.022\\
GIS3&11,949&5.155$\pm$0.025\\
\hline
\end{tabular}

\subsection{GIS Data}

Figure 1 shows the image obtained with the GIS2 detector. Most 
of the detector area is covered by stray light due to the bright 
source GX5--1, located outside the field of view, at an off--axis 
angle of about 40 arcmin. Fortunately, GRS 1758--258 lies in a 
relatively unaffected region of the detector, which allows us to 
estimate the contamination from GX5--1 as explained below.  
The source counts were extracted from a circle of 6 arcmin 
radius centered at the position of GRS 1758--258, and rebinned 
in order to have a minimum of 25 counts in each energy 
channel. Due to the present uncertainties in the ASCA response 
at low energies, we only considered photons in the 0.8--10 keV 
range. The background spectrum was extracted from the 
corresponding regions of observations of empty fields provided 
by the ASCA Guest Observer Facility.  The contribution to the 
background due to GX5--1 is mostly concentrated in a circular 
segment with area $\sim$36 arcmin2 indicated with A in figure 
1. Its spectrum was estimated by the difference of regions A 
and B, and added to the background. A similar procedure was 
followed to extract the GIS3 net spectrum. 
Using XSPEC (Version 9.0) we explored several spectral models 
by simultaneously fitting the data sets of both GIS instruments. 
The best fit was obtained with a power law with photon index 
$1.66\pm 0.03$ and column density $N_H=(1.42\pm 0.04)\times 10^{22}$ cm$^{-2}$ 
(reduced $\chi^2= 1.013$ for 372 d.o.f., errors at 90\% confidence 
intervals for a single interesting parameter). Other models based 
on a single spectral component (e.g. blackbody, thermal 
bremsstrahlung) gave unacceptable results, with the exception 
of the Comptonized disk model of Sunyaev \& Titarchuk (1980). 
However, the limited energy range of the ASCA data alone, does 
not allow in the latter case to pose interesting constraints on the 
fit parameters. 

The GIS instruments have a time resolution of 0.5 s or 62.5 ms, 
according to the available telemetry rate. Most of our data had 
the higher time resolution. Using a Fourier transform technique, 
and after correction of the times of arrivals to the solar system 
barycenter, we performed a search for periodicities. No coherent 
pulsations in the 0.125--1000 s period range were found. For the 
hypothesis of a sinusoidal modulation we can set an upper limit 
of $\sim$5\%  to the pulsed fraction.  

\subsection{SIS Data}

Both SIS instruments were operated in the single chip mode, 
which gives a time resolution of 4 s and images of a square 
11x11 arcmin2 region (Figure 2).  Most of the SIS data (83\%) 
were acquired in "faint mode" and then converted to "bright". 
This allows to minimize the errors due to the echo effects in the 
analog electronics and to the uncertainties in the dark frame 
value (Otani \& Dotani, 1994). The inclusion of the data directly 
acquired in bright mode resulted in spectra of lower quality 
(significant residuals in the 2--3 keV region). We therefore 
concentrated the spectral analysis on the faint mode data. The 
source counts (0.6--10 keV) were extracted from circles with a 
radius of 3 arcmin, and the resulting energy spectra (1024 PI 
energy channels) rebinned in order to have at least 25 counts in 
each bin. We subtracted a background spectrum derived during 
our observation from an apparently source free region of the 
CCDs (see figure 2). This background is higher than that obtained 
from the standard observations of empty fields, probably due to 
the contamination from GX5--1. It contributes $\sim$4\% of the 
extracted counts. We verified that the derived spectral 
parameters do not change significantly if we use the blank sky 
background file, or even if we completely neglect the 
background subtraction.
By fitting together the data from the two SIS we obtained 
results similar to those derived with the GIS instruments. In 
particular, a power law spectrum gives photon index $\alpha=1.70\pm 0.03$ 
and $N_H=(1.55\pm 0.03) \times 10^{22}$ cm$^{-2}$, with a reduced $\chi^2$ of 
1.031 (872 d.o.f.). 
  
No prominent emission lines are visible in the spectrum of GRS 
1758--258 (as already mentioned, some features in the region 
around 2 keV are probably due to instrumental problems, they 
appear stronger when the bright mode data and the 
corresponding response matrix are used). Upper limits on the 
possible presence of an iron line were computed by adding a 
gaussian line centered at 6.4 keV to the best fit power law 
model and varying its parameters (intensity and width) until an 
unacceptable increase in the c2  was obtained.  The 95\% upper 
limit on the equivalent width is $\sim$50 eV for a line width of 
$\sigma=0.1$ keV and increases for wider lines (up to $\sim$110 eV for 
$\sigma=0.5$ keV). 

Also in the case of the SIS, a search for periodicities (limited to 
period greater than 8 s)  resulted only in upper limits similar to 
the GIS ones.

\section{Discussion}

The soft X--ray flux observed with ROSAT in 1993 (Mereghetti et 
al. 1994b) was higher than that expected from the extrapolation 
of the quasi--simultaneous SIGMA measurement ($E>40$ keV), 
indicating the presence of a soft spectral component with power 
law photon index $\sim$3 below $\sim$2 keV. Clearly, such a 
steep, low--energy component is not visible in the present ASCA 
data, which are well described by a single flat power law. The 
corresponding flux of $4.8\times 10^{-10}$ ergs cm$^{-2}$ s$^{-1}$ 
(in the 1--10 keV 
band, corrected for the absorption)  is within the range of values 
measured  in March--April 1990 (Sunyaev et al. 1991), when the 
source was in its highest observed state. This fact is consistent 
with the presence of a prominent soft component only when the 
hard X--ray flux is at a lower intensity level. 

Though a single power law provides an acceptable fit to the 
ASCA data, we also explored spectral models consisting of two 
different components: a soft thermal emission plus a hard tail. 
For instance, with a blackbody plus power law, we obtained a 
good fit to both  the SIS and GIS data with $kT\sim 0.4-0.5$ keV 
and photon index $\sim 1.4-1.5$  ($\chi^2 \simeq 0.98$).  Obviously 
such a power law must steepen at higher energy to be 
consistent with the SIGMA observations. In fact a Sunyaev--Titarchuk 
Comptonization model can equally well fit the ASCA 
hard tail and provide an adequate spectral steepening to match 
the high energy data (see Figure 3). Good results were also 
obtained when the soft thermal component was fitted with 
models of emission from accretion disks (e.g.  Makishima et al. 
1986, Stella \& Rosner 1984). In all cases the total flux in the soft 
component amounts only to a few percent of the overall (0.1--
300 keV) luminosity. However, the low observed flux, coupled 
to the high accretion rates required by the fitted temperatures, 
implies an unplausible large distance for GRS 1758--258 and/or 
very high inclination angles (note that there is no evidence so 
far of eclipses or periodic absorption dips which could hint to a 
high inclination system). A possible alternative solution is to 
invoke a significant dilution of the optically thick soft 
component by Comptonization in a hot corona. A very rough 
estimate shows that, in order to effectively remove photons 
from the thermal distribution, a scattering opacity of 
$\tau_{es}\sim 2-5$ is required. 

Our ASCA observation provides the most accurate measurement 
of the absorption toward GRS 1758--258 obtained so far. 
Obviously the derived value is slightly dependent on the 
adopted spectral model. However, values within at most 10\% of 
$1.5\times 10^{22}$ cm$^{-2}$ were obtained for all the models (one or two 
components)  fitting the data. This column density is consistent 
with a distance of the order of the Galactic center and similar to 
that of other sources in the galactic bulge (Kawai et al. 1988), 
but definitely smaller than that observed with ASCA in 1E 
1740.7--2942 (Churazov et al. 1996). 

The information on the galactic column density, coupled to the 
optical/IR data, can yield some constraints on the possible 
companion star of GRS 1758--258 (see Chen, Gehrels \& Leventhal 
1994).  A candidate counterpart with $I\sim$19 and $K\sim$17 
(Mereghetti et al. 1994a) lies within $\sim$2" of the best radio 
position (Mirabel et al. 1992b). Other infrared sources present in 
the X--ray error circle (10" radius) are fainter than $K\sim 17$  
(Mirabel \& Duc 1992). Using an average relation between $N_H$ 
and optical reddening (Gorenstein 1975), we estimate a value of 
$A_V\sim 7$,  corresponding to less than one magnitude of 
absorption in the K band (Cardelli, Clayton \& Mathis 1989). 
Thus, for a distance of the order of 10 kpc, the K band absolute 
magnitude must be fainter than $M_K\sim 1$. This limit clearly 
rules out supergiant or giant companion stars, as well as main 
sequence stars earlier than type A (Johnson 1966), thus 
excluding the possibility that GRS 1758--258 is in a high mass 
binary system. 
 
The flux of GRS 1758--258 measured with the SIS instruments 
corresponds to a 1--10 keV  luminosity of $4.5\times 10^{36}$ 
ergs s$^{-1}$ (for a distance of 10 kpc).  A reanalysis of archival data from 
TTM/MIR, XRT/Spacelab--2 and EXOSAT (Skinner 1991),  showed 
that GRS 1758--258 had a similar intensity also in 1985 and in 
1989. An earlier discovery had been prevented only by 
confusion problems with GX5--1, much brighter than GRS 1758--
258 below $\sim$20 keV. Subsequent hard X--ray observations 
with SIGMA (Gilfanov et al. 1993, Goldwurm et al. 1994) 
repeatedly detected GRS 1758--258 with a hard spectrum 
extending up to $\sim$300 keV. It is therefore clear that GRS 
1758--258, though variable by a factor of $\sim$10 on a 
timescale of months, is not a transient source.   

\section{Conclusions}

The  ASCA satellite  has provided the first detailed data on GRS 
1758--258  in the 1--10 keV region, allowing to minimize the 
confusion problems caused by the vicinity of GX5--1, that 
affected previous observations with non imaging instruments. 

The possible black hole nature of GRS 1758--258, inferred from 
the high energy data (Sunyaev et al. 1991, Goldwurm et al. 
1994), is supported by the ASCA results. The power law 
spectrum, extending up to the hard X--ray domain is similar to 
that of Cyg X--1 and other black hole candidates in their low (or 
hard) state. Furthermore, our stringent limits on the presence of 
periodic pulsations and accurate measurement of interstellar 
absorption make the possibility of a neutron star accreting from 
a massive companion very unlikely.  The lack of iron emission 
lines in the SIS data has to be confirmed by more stringent 
upper limits to rule out, e.g., the presence of a reflection 
component as proposed for Cyg X--1 (Done et al. 1992).  For 
comparison, the iron line recently observed with ASCA in Cyg X--
1 has an equivalent width of only 10--30 eV (Ebisawa et al. 
1996).

The prominent soft excess observed with ROSAT in 1993, when 
the hard X--ray flux was in a lower intensity state, was absent 
during our observation. The source was in a hard spectral state, 
with a possible soft  component accounting for  $\sim$5\% of the 
total luminosity at most. A similar soft component 
($kT\sim 0.14$ keV), but contributing a larger fraction of the 
flux, has been observed in Cyg X--1 and attributed to emission 
from the accretion disk (Balucinska--Church et al. 1995). If the 
soft component in GRS 1758--258 originates from the disk, 
strong dilution is required. An optically thick hot cloud 
embedding the innermost part of the disk is an attractive 
hypothesis.  To test the viability of this model, a detailed fit to 
simultaneous data over a broad energy range, as available, e.g., 
with SAX in the near future, is required.

\clearpage

\begin{figure}
\caption{Image of GRS 1758--258 obtained with the GIS2 
instrument. The field has a diameter of 50 arcmin. Stray X--
rays from GX5--1, located 40 arcmin north of GRS 1758--258, 
affect mostly part A of the source extraction circle (6' radius).} 
\end{figure}

\begin{figure}
\caption{SIS0 image of GRS 1758--258. The two circles indicate the 
source and background extraction regions, with radii of 3' and   
1.5', respectively.} 
\end{figure}

\begin{figure}
\plotone{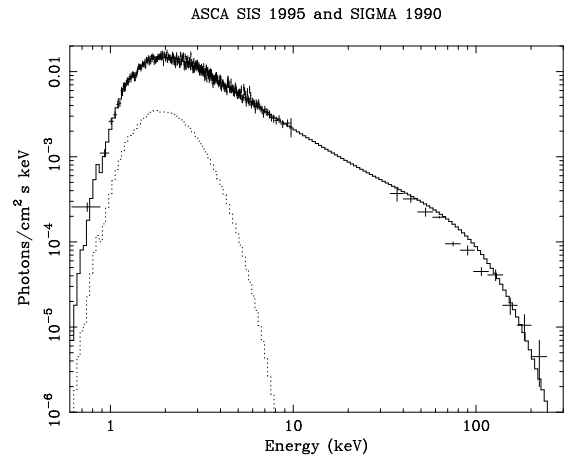}
\caption{The ASCA SIS spectrum of GRS 1758--258 is compared to 
the (non simultaneous) SIGMA data of Spring 1990 (Gilfanov 
et al. 1993).  The solid line correponds to a fit to the ASCA 
data with a blackbody plus Sunyaev \& Titarchuk 
Comptonization model (reduced $\chi^2 0.98$, 870 d.o.f.).  The best 
fit parameters ($\tau_{es}\sim 6$, $kT\sim 25$ keV) are consistent 
with the high energy data. The dotted line shows the soft 
blackbody component ($kT=0.5$ keV).} 
\end{figure}

\end{document}